\documentclass[preprint,preprintnumbers, prd, floatfix, superscriptaddress,
nofootinbib] {revtex4-1}
\usepackage{graphicx}
\usepackage{epsfig}
\usepackage{bm}
\usepackage{amssymb}
\usepackage{float}
\usepackage{amsmath}
\usepackage{dcolumn}
\usepackage{cancel}
\usepackage[colorlinks]{hyperref}
\usepackage[usenames,dvipsnames]{color}
\usepackage{epstopdf}
\hypersetup{
     breaklinks=true,
    pdfstartview={FitH},    colorlinks=true,       
    linkcolor=blue,          
    citecolor=red,        
    filecolor=magenta,      
    urlcolor=blue,           
    anchorcolor=green,      
    linktocpage=true
}

\begin{document}

\title{Matter Power Spectra in Viable $f(R)$ Gravity Models with
Dynamical Background}

\author{Yow-Chun Chen}
\affiliation{Chongqing University of Posts \& Telecommunications, Chongqing 400065 \\
Department of Physics, National Tsing Hua University and \\
National Center for Theoretical Sciences, Hsinchu 300}

 \author{Chao-Qiang Geng}
\affiliation{Chongqing University of Posts \& Telecommunications, Chongqing 400065 \\
Department of Physics, National Tsing Hua University and \\
National Center for Theoretical Sciences, Hsinchu 300}

\author{Chung-Chi Lee}
\affiliation{
DAMTP, Centre for Mathematical Sciences, University of Cambridge, Wilberforce Road, Cambridge CB3 0WA}

\author{Hongwei Yu}
\email{hwyu@hunnu.edu.cn}
\affiliation{Synergetic Innovation Center for Quantum Effects and Applications,\\
and Department of Physics, Hunan Normal University, Changsha, Hunan 410081}

\begin{abstract}
We study the matter power spectra in  the viable $f(R)$ gravity models with the dynamical background evolution and  linear perturbation theory by using the CosmoMC package. We show that  these viable $f(R)$ models generally shorten the age of the universe and suppress the matter density fluctuation. We examine the allowed ranges of the model parameters and the constraints of the cosmological variables from the current observational data, and find that the dynamical evolution of $\rho_{DE}(z)$ plays an important role to constrain the neutrino masses.
\end{abstract}

\maketitle

\section{Introduction}\label{sec:1}

According to the recent cosmological observations,  our universe is undergoing a late-time accelerating expansion phase, 
which can be realized by introducing a time independent vacuum energy, called dark energy, built in the
$\Lambda$CDM model~\cite{Amendola:2015ksp}.
Although this standard model of cosmology fits well with the observational data, it fails to solve the cosmological constant problem, 
related to the hierarchy~\cite{Weinberg:1988cp, WBook} and coincidence~\cite{Peebles:2002gy, Ostriker:1995rn, ArkaniHamed:2000tc}
 ones. These problems have motivated people to explore new theories beyond $\Lambda$CDM, such as
 those with the dynamical dark energy~\cite{Copeland:2006wr}. A typical model of such theories 
is to modify the standard general relativity (GR) by promoting the Ricci scalar of $R$ 
in the Einstein-Hilbert action to an arbitrary function,   i.e., $f(R)$~\cite{DeFelice:2010aj}.

In this study, we concentrate on the viable $f(R)$ gravity theories, which satisfy
several  conditions based on  the theoretical and observational constraints~\cite{DeFelice:2010aj}.
Some of the typical viable $f(R)$ gravity models have been discussed to fit the cosmological evolutions~\cite{Yang:2010xq}.
It has been shown that
the viable $f(R)$ gravity  can perfectly describe the power spectrum of the matter density fluctuation~\cite{Hu:2013twa, Raveri:2014cka, deMartino:2015zsa, Geng:2014yoa, Geng:2015vsa} and the formation of the large scale structure (LSS)~\cite{Li:2011vk, Puchwein:2013lza, Lombriser:2013wta, Llinares:2013jza}.
To investigate the dynamical dark energy models, one needs to use the existing  open-source programs.
However, most of these programs are written with either the parametrization in term of the equation of state
or the background evolution being the same as the $\Lambda$CDM model as used in Refs.~\cite{Geng:2014yoa, Geng:2015vsa}.
In this analysis, we would explore the allowed parameter spaces of the cosmological observables in the viable $f(R)$ gravity models
when the dynamical background evolution is taken into account. 
In particular, we will show that the allowed windows for the active neutrino masses are further constrained, indicating that the dynamical background 
is indeed important in these viable $f(R)$  theories.

One of the useful program to examine the viable $f(R)$ models is the Modification of Growth with Code for Anisotropies in the Microwave Background (MGCAMB)~\cite{Lewis:1999bs, Hojjati:2011ix}, in which the growth equations of the scalar perturbations and  density fluctuations in the Newtonian gauge are modified.
In addition, we take the dynamical background evolution of  dark energy and test these viable models by using the Cosmological MonteCarlo (CosmoMC)~\cite{Lewis:2002ah} program together with the latest cosmological observational data.

The paper is organized as follows. In Sec.~\ref{sec:2}, we review some basic concepts of the viable  $f(R)$ gravity models.
In particular, we include the perturbation equations of the background evolution of these models. 
In Sec.~\ref{sec:3}, we present our numerical results for the model parameters as well as constraints on the cosmological 
variables in the  models. We give the conclusion  in Sec.~\ref{sec:4}.

\section{Viable $f(R)$ Gravity}\label{sec:2}
\subsection{Viable $f(R)$ models}

The action of $f(R)$ gravity is to extend the Ricci scalar of $R$ in the Einstein-Hilbert action to an arbitrary function, given by
\begin{equation}
\label{eq:action}
S=\int{ d^{4}x \frac{\sqrt{-g}}{2\kappa^2} f(R)}+S_{M}
\end{equation}
where $\kappa^{2}=8\pi G$ with $G$  the Newton's constant, $g$ is the determinant of the metric tensor $g_{\mu\nu}$ and $S_{M}$ is the action of the relativistic and non-relativistic matter.
Varying the action of Eq.~\eqref{eq:action} with respect to $g_{\mu\nu}$, we derive the modified Einstein equation~\cite{Sotiriou:2008rp},
\begin{equation}
\label{eq:field}
f_R R_{\mu\nu}-\frac{f}{2}g_{\mu\nu}-\left(\nabla_{\mu}\nabla_{\nu}-g_{\mu\nu}\Box\right) f_R = \kappa^{2}T^{\left(M\right)}_{\mu\nu}
\end{equation}
where 
$f_R \equiv df(R)/dR$, $\nabla_{\mu}$ is the covariant derivative, $\Box \equiv g^{\mu \nu}\nabla_{\mu} \nabla_{\nu}$ is the d'Alembert operator, and $T^{\left(M\right)}_{\mu\nu}$ is the energy momentum tensor.

To describe the late-time dark energy problem,  $f(R)$ gravity has  to satisfy several viable 
conditions~\cite{Amendola:2015ksp,Amendola:2006we,Bamba:2010iy}, 
including 
(i) a positive effective gravitational coupling, leading to $f_R > 0$;
(ii) a stable cosmological perturbation and a positivity of the  gravitational wave for the scalar mode, causing to $f_{RR} > 0$;
(iii) an asymptotic behavior to the $\Lambda$CDM model in the large curvature region, i.e., $f(R) \rightarrow R - 2 \Lambda$ at $z \gg 0$;
(iv) a late-time stable de-Sitter solution; and
(v) a suitable chameleon mechanism that makes  $f(R)$ gravity passing the local system constraints.
The models who meet with these five conditions  are so-called viable $f(R)$ gravity models.
Note that, due to the viable condition (iv), the viable $f(R)$ models approach to the $\Lambda$CDM model in the high redshift regime, 
allowing us to classify these  models into two classes: the power law  and  exponential types~\cite{Lee:2012dk},
corresponding to 
the two popular viable models:
the Starobinsky~\cite{Starobinsky:2007hu} and  exponential~\cite{Tsujikawa:2007xu, Zhang:2005vt, Cognola:2007zu, Linder:2009jz}
gravity ones, which have the explicit forms,
\begin{eqnarray}
\label{eq:star}
f^{s} (R) &=& R - \lambda_s R_{ch} \left[ 1- \left(1+\frac{R^2}{R_{ch}^2} \right)^{-n} \right] \,, \\
\label{eq:exp}
f^{e} (R) &=& R - \lambda_e R_{ch}\left(1-e^{-R/R_{ch}} \right) \,,
\end{eqnarray}
respectively, where $\lambda_i$ and $n$ are the dimensionless model parameters, and $R_{ch}$ is the constant characteristic curvature in each model.
At the high $z$ region, we can see that  $f(R) \rightarrow R - \lambda R_{ch}$, which
 plays the role of the cosmological constant and has the same order of magnitude as 
the dark energy density at the present time, leading to $R_{ch} \sim \Lambda_{obs} / \lambda$, where $\Lambda_{obs} = \kappa^2 \rho_{DE}^{(0)}$ is defined from the current value of the dark energy density.
As a result,  a smaller $R_{ch}$ is required for a bigger $\lambda$
to fit the observations, resulting in a relatively larger value of $R/R_{ch}$.
Clearly,  $f(R)$ gravity approaches the $\Lambda$CDM model when $\lambda \gg 1$ and $R_{ch} \rightarrow 0$.

\subsection{Cosmological evolution}
\label{sec:2b}
We describe the universe by using the Friedmann-Lemaitre-Robertson-Walker (FLRW) metric,
\begin{equation}
\label{nonmetric}
ds^{2} = g_{\mu\nu}dx^{\mu}dx^{\nu} = -dt^{2}+a^{2} \left( t \right) d\vec{x}^2
\end{equation}
where $a(t)$ is the scale factor.
By introducing this metric into the modified Einstein field equation in Eq.~\eqref{eq:field}, we obtain the modified Friedmann equations,
\begin{eqnarray}
\label{firstFriedmann}
&& 3 f_R H^{2}=\frac{1}{2}\left(f_R R-f\right)-3H\dot{f_R}+\kappa^{2}\rho_{M} \,, \\
\label{secondFriedmann}
&& 2 f_R \dot{H} = -\ddot{f_R} + H\dot{f_R} - \kappa^{2}\left(\rho_{M}+P_{M}\right) \,,
\end{eqnarray}
where the dot represents the derivative with respect to the cosmic time $t$, $H\equiv \dot{a} / a$ is the Hubble parameter, 
and $\rho_M = \rho_r + \rho_m$ ($P_M = P_r + P_m$) is the energy density (pressure) of relativistic ($r$) and non-relativistic ($m$) fluids.
Comparing to the original Friedmann equations of
\begin{eqnarray}
3 H^2 = \kappa^2 \left( \rho_M + \rho_{DE} \right) \quad \mathrm{and} \quad 2\dot{H} = - \kappa^2 \left( \rho_M + P_M +\rho_{DE} + P_{DE} \right) \,,
\end{eqnarray}
the effective energy density and pressure of dark energy can be defined by
\begin{eqnarray}
&& \kappa^2 \rho_{DE} = \frac{1}{2}\left(f_R R-f\left(R\right)\right)-3H\dot{f_R}+3\left(1-f_R\right)H^2 \,, \\
&& \kappa^2 P_{DE} = -\frac{1}{2}\left(f_R R-f(R)\right)+\ddot{f_R}+2H\dot{f_R}-\left(1-f_R\right)\left(2\dot{H}+3H^2\right) \,,
\end{eqnarray}
respectively, 
which are easily checked to satisfy the continuity equation,
\begin{eqnarray}
\label{continuity}
\dot{\rho}_{DE}+3H\left(\rho_{DE}+P_{DE}\right)=0 \,.
\end{eqnarray}

\begin{figure}[htbp]
\centering
\includegraphics[width=0.9 \linewidth]{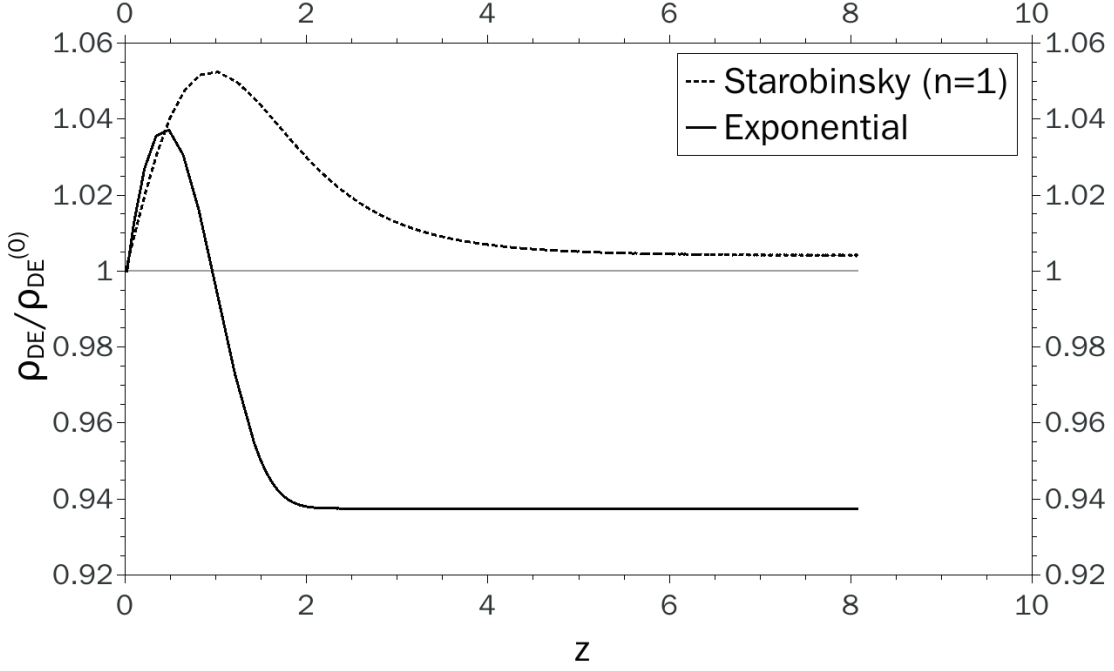}
\vskip 0 pt
\caption{ $\rho_{DE}/\rho_{DE}^{(0)}$ as  functions of the redshift $z$ with $\rho_{DE}^{(0)} \equiv \rho_{DE}(z=0)$,
where the solid and dashed lines represent the exponential  and Starobinsky gravity models with $\lambda_e^{-1} = 0.56 $ and $(\lambda_s^{-1}, n) = (0.28, 1)$, respectively,
while $H_0=67.8$ $(km/s \cdot Mpc)$ and the initial condition is $(\Omega_m, \Omega_r, \Omega_{DE} ) = (0.306, 7.88 \times 10^{-5}, 0.694) $.}
\label{fig:1}
\end{figure}


Following the same processes in Refs.~\cite{Hu:2007nk, Bamba:2010ws}, Eqs.~\eqref{firstFriedmann} and \eqref{secondFriedmann} 
can be reduced to a second order differential equation,
\begin{eqnarray}
\label{diffeqyh}
y_{H}^{\prime \prime} + J_{1} y_{H}^{\prime} + J_{2}y_{H} + J_{3} = 0 \,,
\end{eqnarray}
with
\begin{eqnarray}
\label{yhyr}
y_{H}\equiv \frac{\rho_{DE}}{\rho_{m}^{(0)}}=\frac{H^{2}}{m^{2}}-a^{-3}- \chi_r a^{-4}\,, \quad  
y_{R}\equiv\frac{R}{m^{2}} -3a^{-3} 
\end{eqnarray}
and 
\begin{eqnarray}
\label{j1}
&& J_{1}=4+\frac{1}{y_{H}+a^{-3}+\chi a^{-4}}\frac{1 - f_R}{6m^{2}f_{RR}} \,, \\
\label{j2}
&& J_{2}=\frac{1}{y_{H}+a^{-3}+\chi a^{-4}}\frac{2-f_R}{3 m^{2}f_{RR}} \,, \\
\label{j3}
&& J_{3}=-3a^{-3}-\frac{\left( 1 - f_R \right)\left(a^{-3}+2\chi a^{-4}\right)+\left(R-f\right)/3 m^{2}}{y_{H}+a^{-3}+\chi a^{-4}}\frac{1}{6 m^{2}f_{RR}} \,,
\end{eqnarray}
where $m^{2}\equiv \kappa^2 \rho_{m}^{(0)}/3$, $\rho_i^{(0)} \equiv \rho_i (z=0)$, and $\chi_r \equiv \rho_{r}^{(0)} / \rho_{m}^{(0)}$  is the energy density ratio between relativistic and non-relativistic fluids at present time.
Here,   the prime in Eq.~(\ref{diffeqyh}) denotes the derivative with respect to $\ln a$.

In Fig.~\ref{fig:1}, we illustrate the evolutions of the dark energy densities as  functions of the redshift $z$ by solving Eq.~\eqref{diffeqyh} with $H_0=67.8$ $(km/s \cdot Mpc)$ and the initial condition of $(\Omega_m, \Omega_r, \Omega_{DE} ) = (0.306, 7.88 \times 10^{-5}, 0.694) $,
where $\lambda_e^{-1} = 0.56 $  and $(\lambda_s^{-1}, n) = (0.28, 1)$, corresponding to the exponential  (solid line) and Starobinsky (dashed line) models, are taking  from the central values in Table.~\ref{table2},  respectively.
One can observe that the evolutions of $\rho_{DE}$ for the two viable $f(R)$ models are frozen in the high redshift regime, 
and start to evolve at the late-time of the universe.
 The dynamical behaviors become important for $z \lesssim 4$ and $2$ in the Starobinsky and exponential models, respectively.
The dark energy density increases in time at the beginning and decreases when the redshift close to zero, {\em i.e.} $z \to 0$, 
leading to $\rho_{DE}(z \lesssim 0.5) \geq \rho_{DE}^{(0)}$, which covers the dark energy dominated epoch.
As a result, we obtain
\begin{eqnarray}
\label{eq:age_l}
&& t_{age}^{\Lambda CDM} = 13.83 \ (Gyrs) \qquad \mathrm{(\Lambda CDM \ model)} \\
\label{eq:age_s}
&& t_{age}^{s} = 13.78 \ (Gyrs) \qquad \mathrm{(Starobinsky \ model)} \\
\label{eq:age_e}
&& t_{age}^{e} = 13.80 \ (Gyrs) \qquad \mathrm{(exponential \ model)},
\end{eqnarray}
 where we have used 
\begin{eqnarray}
t_{age} = \frac{1}{H_0} \int_0^1 \frac{da}{a \sqrt{\Omega_r a^{-4} + \Omega_m a^{-3} + \Omega_{DE}(a)}}
\end{eqnarray}
as the age of the universe.
Note that the bigger $t_{age}$ is, the longer time the LSS of the universe has to grow.
From Eqs.~\eqref{eq:age_l}-\eqref{eq:age_e}, one finds that $t_{age}^{s} < t_{age}^{e} < t_{age}^{\Lambda CDM}$, 
so that the dynamical background evolutions in the viable $f(R)$ gravity theories suppress the matter density fluctuations.

\subsection{Linear perturbation theory}
\label{sec:2c}

We review the linear perturbation theory with two scalar-mode perturbations, $\Psi$ and $\Phi$.
Following the similar processes in Refs.~\cite{Tsujikawa:2007gd, Ma:1995ey}, the metric in the Newtonian gauge is
\begin{eqnarray}
\label{eq:metric}
ds^{2}=-\left(1+2\Psi\right)dt^{2}+a\left(t\right)\left(1-2\Phi\right)d\vec{x}^{2} \,,
\end{eqnarray}
Substituting Eq.~\eqref{eq:metric} into Eq.~\eqref{eq:field}, the perturbation equations in the $k$ space are given by,
\begin{eqnarray}
\label{eq:firstperturb}
\frac{k^{2}}{a^{2}}\Phi
&=& -3H\left(\dot{\Phi}+H\Psi\right)-\frac{1}{2 f_R}\left[-3H\dot{\delta f_R}+\left(3H^{2}+3\dot{H}-\frac{k^{2}}{a^{2}}\right) f_{RR} \delta R\right]
\nonumber \\
&& -\frac{1}{2 f_R}\left[3 \dot{f_R}\left(\dot{\Phi}+H\Psi\right)+3H\dot{f_R}\Psi+\kappa^{2}\delta\rho_{M}\right] \,, \\
\label{eq:secondperturb}
\Phi &=& \Psi+\frac{f_{RR}}{f_R} \delta R \,,
\end{eqnarray}
resulting in
\begin{eqnarray}
\label{eq:reducedperturb}
 \frac{k^2}{a^2}\Psi = -4\pi G \mu\left(k,a\right)\rho_M \Delta_M  \quad \mathrm{and} \quad \frac{\Phi}{\Psi} = \gamma\left(k,a\right)
\end{eqnarray}
with
\begin{eqnarray}
\label{eq:parmugamma}
\mu\left(k,a\right)=\frac{1}{f_R}\frac{1+4\frac{k^2}{a^2}\frac{f_{RR}}{f_R}}{1+3\frac{k^2}{a^2}\frac{f_{RR}}{f_R}} \quad \mathrm{and} \quad \gamma\left(k,a\right)=\frac{1+2\frac{k^2}{a^2}\frac{f_{RR}}{f_R}}{1+4\frac{k^2}{a^2}\frac{f_{RR}}{f_R}} \,,
\end{eqnarray}
where $k$ is the comoving wavenumber and $\Delta_M \equiv \delta_{M}+3 H\left(1+\omega_M\right)v_M/k$ is the gauge-invariant matter density perturbation with $w_M = P_M / \rho_M$  the equation of state 
 and $v_M$  the velocity for matter.
The growth equation for the matter density perturbation at the matter dominated epoch can be derived from Eqs.~\eqref{eq:firstperturb} and \eqref{eq:secondperturb} with $v_m = 0$, given by
\begin{eqnarray}
\label{eq:deltam}
\ddot{\delta}_m + 2H\dot{\delta}_m - 4 \pi G \mu(k,a) \rho_m \delta_m = 0 \,.
\end{eqnarray}
Both the Starobinsky and  exponential $f(R)$ models have the conditions of $0 < f_R < 1$ and $f_{RR} > 0$, implying  that (i) the larger $k$ is, the bigger $\mu(k,a)$ behaves, and (ii) $\mu(k,a) \simeq f_R^{-1} > 1$ at the super-horizon scale with $k \rightarrow 0$, indicating that the matter density fluctuations are enhanced 
due to
 the scale independent and dependent factors of
 $f_R^{-1}$ and $\left( 1+4\frac{k^2}{a^2}\frac{f_{RR}}{f_R}\right) / \left(1+3\frac{k^2}{a^2}\frac{f_{RR}}{f_R} \right)$, respectively,
 the these models.

To simplify the calculations,  the background evolution of the dark energy density is usually taken to be the same as that in 
the $\Lambda$CDM model, so that the Hubble parameter in Eq.~(\ref{eq:deltam}) is approximately given by
\begin{eqnarray}
H \simeq H_0 \sqrt{\Omega_r a^{-4} + \Omega_m a^{-3} + \Omega_{DE}^{(0)}}
\end{eqnarray}
with $\Omega_{DE}^{(0)} = \Omega_{DE}(z = 0)$, the constant $\rho_{DE}$.
It is clear that the $f(R)$ gravity theory becomes the $\Lambda$CDM limit of $f(R) = R - 2 \Lambda$ at $z\gg 0$, corresponding to $\mu = \gamma = 1$.
One can further see from Eq.~\eqref{eq:deltam} that the scale dependent $\mu(k,a)$ is bigger when $k$ is larger, indicating that the growth of $\delta_m$ at the sub-horizon scale is faster than that at the super-horizon.
On the other hand, the massive neutrino suppresses the matter density fluctuation at the scale smaller than the free-streaming length~\cite{Lesgourgues:2006nd}, which  is opposite to the $f(R)$ enhancement.
Thus, the allowed windows for  the neutrino masses sum, $\Sigma m_{\nu}$, in these $f(R)$ gravity models can be broader than that in the $\Lambda$CDM one~\cite{Motohashi:2010sj, Motohashi:2012wc}.

In Refs.~\cite{Geng:2014yoa, Motohashi:2010sj, Motohashi:2012wc}, the matter power spectrum of $P(k)$ 
 and the active neutrino masses were explored in the viable $f(R)$ gravity models with the $\Lambda$CDM background evolution. 
However, with the dynamical background, 
the Hubble parameter in Eq.~\eqref{eq:deltam} should be replaced by
\begin{eqnarray}
H = H_0 \sqrt{\Omega_r a^{-4} + \Omega_m a^{-3} + \Omega_{DE}(z)} \,,
\end{eqnarray}
and $t_{age}$ is shortened as well.
Clearly, the behavior for $P(k)$ and the constraint for $\Sigma m_{\nu}$ with $\rho_{DE}(z) \neq constant$ would be slightly different from those with the $\Lambda$CDM background one.
We modify the dynamical background evolution $\rho_{DE}(z)$ in the MGCAMB program to study the difference between the viable $f(R)$ and  $\Lambda$CDM models.
To achieve this goal, we calculate a lookup table for $\rho_{DE}(z)$, which is the evolution of the dark energy density with the boundary condition $\Omega_{DE}^{(0)} = \rho_{DE}(z=0)/\rho_c$.
By using the interpolation method, the MGCAMB program is able to use the redshift dependent dark energy density, 
in which the Hubble parameter is evaluated from
\begin{eqnarray}
H^2 =\frac{ \kappa^2 \rho(z) }{3}= \frac{\kappa^2}{3} \left[  \rho_c \left( \Omega_m a^{-3} + \Omega_r a^{-4} \right) + \rho_{DE}(z) \right] \,.
\end{eqnarray}
Subsequently,  $\delta_{m,r}$ can be obtained with the dynamical background evolution.

\begin{figure}[htbp]
\centering
\includegraphics[angle=-90,width=0.9 \linewidth]{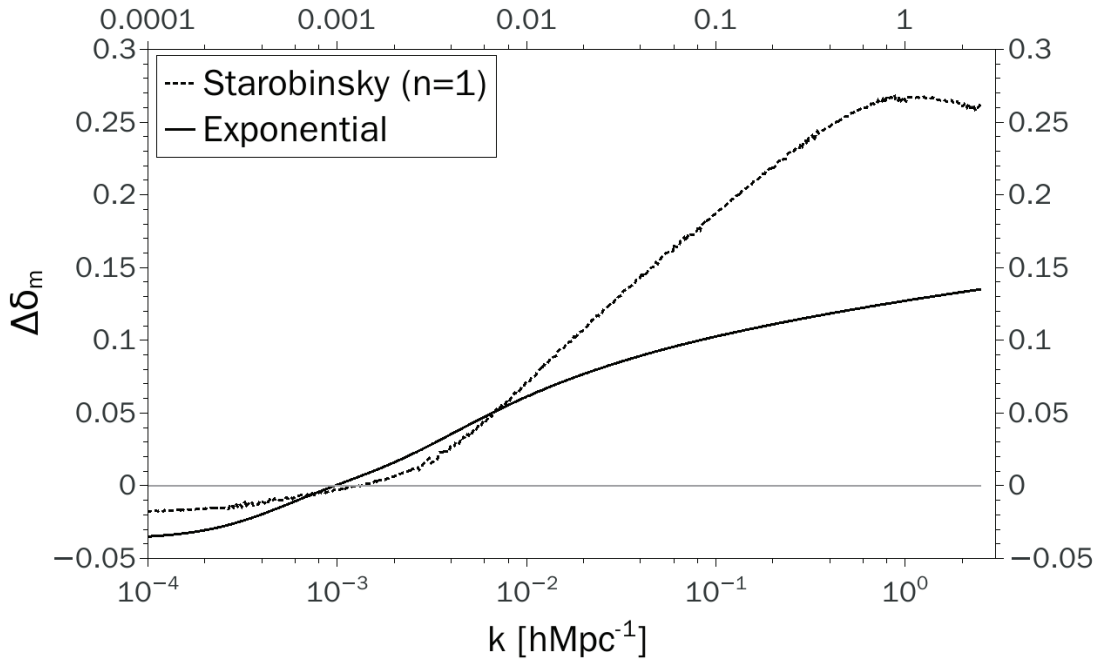}
\vskip 0 pt
\caption{$\Delta\delta_{m}\equiv(\delta^{f(R)}_m / \delta^{\Lambda CDM}_{m})-1$ as  functions of $k$, 
where $\delta^{f(R)}_m$ and $\delta^{\Lambda CDM}_{m}$ are the matter densities of perturbations in the $f(R)$ gravity models
with the dynamical background and that in the $\Lambda$CDM model, respectively,
while the legends are the same as those in Fig.~\ref{fig:1} 
}
\label{fig:4}
\end{figure}

The growth of $\delta_m$ in the $f(R)$ gravity model is suppressed due to the shorter $t_{age}$, but enhanced at the linear perturbation level.
To estimate which effect plays a more important role in the evolution history, the numerical calculation is used.
In Fig.~\ref{fig:4},
we present $ \Delta \delta_m \equiv (\delta^{f(R)}_m / \delta^{\Lambda CDM}_{m})-1$ as  functions of $k$,
where $\delta^{f(R)}_m$ and $\delta^{\Lambda CDM}_{m}$ are the matter densities of perturbations in the $f(R)$ gravity models with the dynamical background and that in the $\Lambda$CDM one.
To plot this figure, we choose the same primordial matter density fluctuation for the Starobinsky and  exponential models as well as the  $\Lambda$CDM one, 
i.e., $\delta^{\Lambda CDM}_m(z \gg 0)=\delta^{f(R)}_m(z \gg 0)$.
The other boundary conditions and  model parameters are the same as those in Fig.~\ref{fig:1}, while
the solid and dashed lines represent the results with the exponential and Starobinsky gravity models, respectively.
We can see that the shorter value of $t_{age}$ in  $f(R)$ gravity is more important at the super-horizon scale, 
leading to $\Delta \delta_m < 0$ for $k \lesssim 10^{-3}$, 
whereas the $f(R)$ effect overcomes the suppression and $\Delta \delta_m$ turns into positive at a large $k$.
In addition, from Eqs.~\eqref{eq:age_l} - \eqref{eq:age_e}, we can estimate that the suppression for the Starobinsky model should be more significant 
than that for the exponential one, but this behavior is flipped by the scale independent effect from  $f(R)$ gravity, resulting in $\Delta \delta_m^{s} < \Delta \delta_m^{e} < 0$ when $k \rightarrow 0$.

\section{Constraints from Cosmological Observations}
\label{sec:3}
\begin{center}
\begin{figure}[htbp]
\centering
\includegraphics[width=0.8 \linewidth]{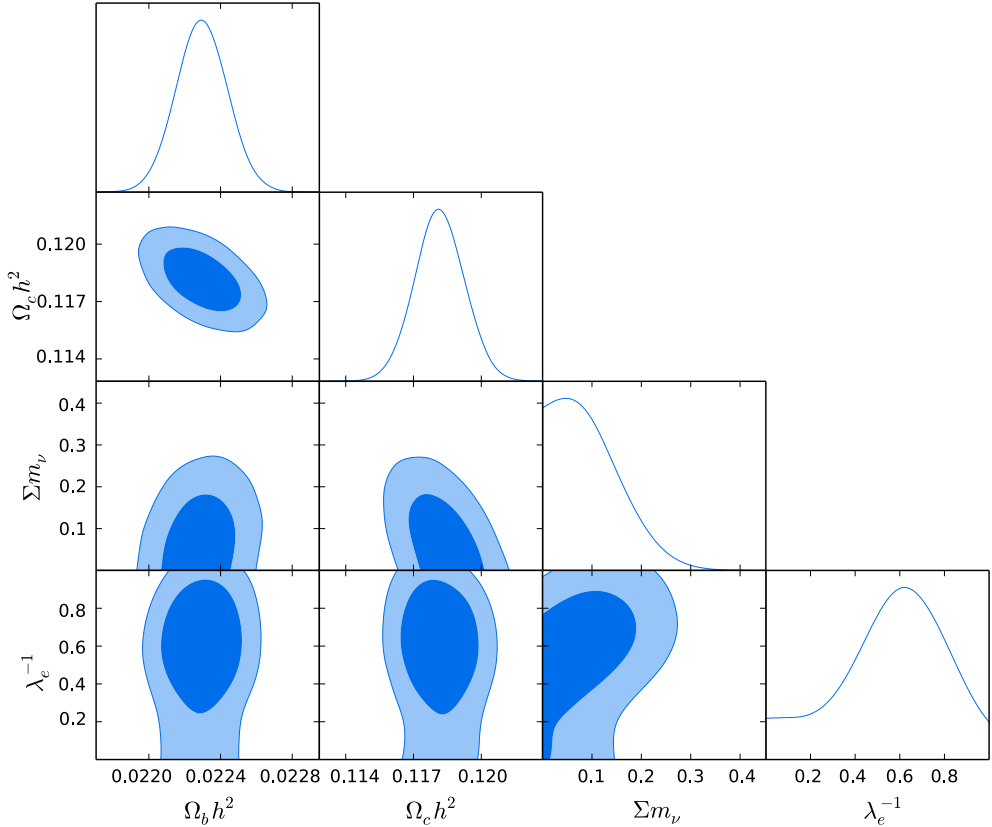}
\caption{Two-dimensional contour plots of $\Omega_b$, $\Omega_c$, $\Sigma m_{\nu}$ and $\lambda_e^{-1}$ in the exponential gravity model.}
\label{fig:2}
\end{figure}
\end{center}

\begin{center}
\begin{figure}[htbp]
\centering
\includegraphics[width=0.8 \linewidth]{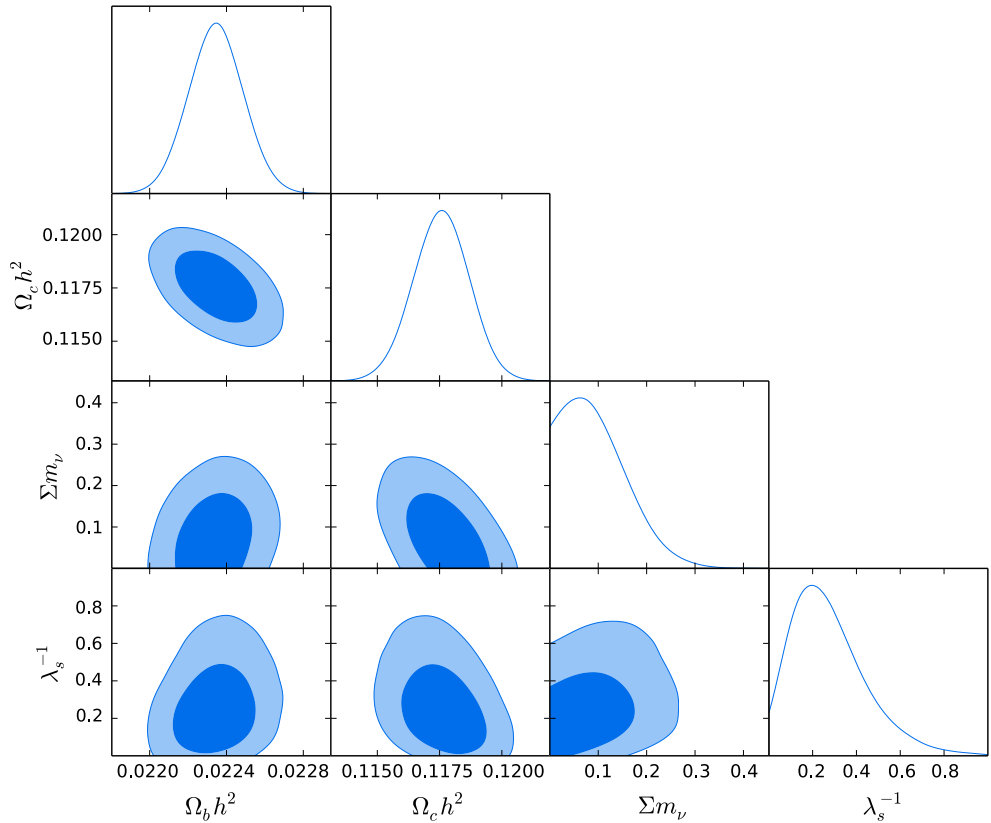}
\caption{Legend is the same as Fig.~\ref{fig:2} but in 
the Starobinsky gravity model with $n=1$.}
\label{fig:3}
\end{figure}
\end{center}

We merge our version of MGCAMB into the CosmoMC
 package to extract the allowed windows for the cosmological parameters together with the latest cosmological observational data, including the Cosmic Microwave Background radiation (CMB) from {\it Planck 2015}~\cite{Ade:2015xua}, Baryon Acoustic Oscillations (BAO) from Baryon Oscillation Spectroscopic Survey (BOSS)~\cite{Anderson:2013zyy}, and type Ia supernova from Supernova Legacy Survey (SNLS)~\cite{Astier:2005qq}.
The prior is listed in Table~\ref{table1}, and our fitting results are shown in Table~\ref{table2}.
We illustrate the contour plots in the exponential and Starobinsky gravity models  in Figs.~\ref{fig:2} and \ref{fig:3}, respectively.
The $\Lambda$CDM limit can be recovered at $\lambda_i^{-1} \to 0$ in the both models.
One finds that the central values of $\lambda_e^{-1} = 0.566 $ and $\lambda_s^{-1} = 0.282$, deviating from zero, 
and the best fitted $\chi^2$ values of $\chi^2_\mathrm{best-fit} = 13458.85$ and $13458.76$ 
in the exponential and Starobinsky  models are both slightly smaller than that in the $\Lambda$CDM one.
It is interesting to note that the corresponding characteristic curvature for the exponential (Starobinsky) model is given  by
$R_{ch} = 6 r \lambda^{-1}_{e} \Omega_{DE} H_0^2 \simeq 2.206 H_0^2$ ($1.192 H_0^2$) with $\Omega_{DE} = 0.694$ and
$r=\rho_{DE}(z \gg 0)/\rho_{DE}(z = 0)
 = 0.936 (1.015)$, as shown in Fig.~\ref{fig:1}.
These results imply that the viable $f(R)$ models are good enough in describing the evolution of the universe.
\begin{table}[htbp]
\begin{center}
\caption{ List of the prior for the cosmological parameters, where $\theta_{MC}$ is the sound horizon to the angular diameter distance, and $A_s$ is the primordial superhorizon power in the curvature perturbation on 0.05 $Mpc^{-1}$ scales.}
\begin{tabular}{|c||c|} \hline
Parameters & Priors
\\ \hline
Model parameter & $10^{-4}< \lambda_i^{-1} <1$
\\ \hline
Baryon density & $5 \times 10^{-3}<\Omega_bh^2<0.1$
\\ \hline
CDM density & $10^{-3}<\Omega_ch^2<0.99$
\\ \hline
Neutrino mass & $0<\Sigma m_{\nu} < 1 $ eV
\\ \hline
Spectral index & $ 0.8 < n_s < 1.2$
\\ \hline
Scalar power spectrum amplitude & $ 2 <\mathrm{ln}(10^{10} A_s) < 4$
\\ \hline
Reionization optical depth & $ 0.01 <\tau <0.8$
\\ \hline
$100 \ \theta_{MC}$  & $ 0.5 <100 \ \theta_{MC} <10$
\\ \hline
Hubble parameter $(km/s \cdot Mpc)$  & $ 20 < H_0 < 100$
\\ \hline
\end{tabular}
\label{table1}
\end{center}
\end{table}

\begin{table}[htbp]
\begin{center}
\caption{Our results within $95\%$ confidence level (C.L.) in the $\Lambda$CDM and viable $f(R)$ gravity models, 
where $\lambda_i^{-1} (i=e,s)$
with 68\% C.L. are the inverses of the model parameters in the exponential  and Starobinsky models, respectively.}
\begin{tabular}{|c||c|c|c|} \hline
\  & Exponential & Starobinsky & $\Lambda$CDM
\\ \hline
$\Sigma m_{\nu}$~(eV) & $ < 0.221$ & $ <0.221$ & $ < 0.202$
\\ \hline
$100\Omega_b h^2$ & $2.23 \pm 0.03 $ & $2.23^{+0.03}_{-0.02}$ & $ 2.23 \pm 0.03 $
\\ \hline
$\Omega_c h^2$ & $0.118 \pm 0.002$ & $0.118 \pm 0.003$ & $0.118 \pm 0.002$
\\ \hline
$n_s$ & $0.969 \pm 0.007$ & $0.970^{+0.008}_{-0.007}$ & $ 0.969^{+0.008}_{-0.007} $
\\ \hline
$\mathrm{ln}(10^{10} A_s)$ & $ 3.047^{+0.057}_{-0.059} $ & $3.026 \pm 0.061$ & $3.071^{+0.054}_{-0.052}$ 
\\ \hline
$\tau$ & $0.059^{+0.030}_{-0.031}$ & $0.049^{+0.032}_{-0.031}$ & $ 0.071 \pm 0.028 $
\\ \hline
$100 \ \theta_{MC}$  & $1.0409 \pm 0.0006 $ & $1.0410 \pm 0.0006 $ & $ 1.0409 \pm 0.0006 $
\\ \hline
$\sigma_8$ & $ 0.877^{+0.046}_{-0.065} $ & $0.942^{+0.040}_{-0.044}$ & $0.811^{+0.025}_{-0.026}$ 
\\ \hline
$H_0$ & $ 67.32^{+1.38}_{-1.40} $ & $67.22^{+1.45}_{-1.91}$ & $67.69^{+1.14}_{-1.24}$ 
\\ \hline
$t_{age} / Gyrs$ & $ 13.80^{+0.08}_{-0.06} $ & $13.81^{+0.07}_{-0.06}$ & $13.81^{+0.07}_{-0.06}$ 
\\ \hline
$\lambda_i$ ($68 \%$ C.L.) & $0.566^{+0.285}_{-0.174}$ & $0.282^{+0.099}_{-0.216}$ & $-$
\\ \hline
$\chi^2_\mathrm{best-fit}$ & $13458.85$ & $13458.76$ & $13459.29$
\\ \hline
\end{tabular}
\label{table2}
\end{center}
\end{table}

 It is well known that the neutrino masses play an important role in both cosmology and  particle physics.
From Table~\ref{table2}, our numerical results show that the allowed regions for $\Sigma m_{\nu}$ in both the exponential and  Starobinsky models are 10\% released from the $\Lambda$CDM model, which are a little different from the conclusions in Refs.~\cite{Geng:2014yoa, Geng:2015vsa}, in which the power-law type model has the capacity to have a larger neutrino mass sum.
This phenomenon may come from the inclusion of the dynamical background evolution, which shortens the age of the universe and suppresses the growth of $\delta_m$.
Furthermore, from Fig.~\ref{fig:4} we can expect that the enhancement of the matter density fluctuation raises $\sigma_8$, especially for the Starobinsky model, so that the primordial power $A_s$ is lowered down in order to fit the observational data, $A_s^{star} \lvert_ {best-fit} \lesssim A_s^{exp} \lvert_ {best-fit}  \lesssim A_s^{\Lambda CDM} \lvert_ {best-fit} $.
In addition, the data prefers the age of the universe to be 13.8 $Gyrs$, even if the dynamical background evolution is included, which means that the best fit of $H_0$ is reduced in the $f(R)$ models.
The effects on $t_{age}$ are cancelled out with each other between the larger value of $H_0$ and the dynamical background evolution.
As a result, the dynamical evolution of $\rho_{DE}$ in $f(R)$ gravity plays a significant role at the linear perturbation level.
Clearly, one should not ignore this when  investigating the CMB polarization and  LSS formation.

\section{Conclusions}\label{sec:4}

We have chosen the exponential and  Starobinsky gravity models to represent the exponential and power-law types of the viable
$f(R)$ models 
and explored the effects from the dynamical background evolution when we perform the observational constraints on these models.
We have shown that the best-fitted  $\chi^2$ values in both models are slightly less than that in the $\Lambda$CDM model.
Even though the linear perturbation theory of  $f(R)$ gravity enhances the matter power spectrum at the small scale by modifying the effective Newtonian constant, the dynamical evolution of $\rho_{DE}$ gives a significant contribution at the low redshift regime and shortens the age of the universe, which equivalently suppresses $\delta_m$ at the small scale.
Clearly, the $f(R)$ enhancement at the small scale is partially compensated when the time-dependent $\rho_{DE}$ is taken into account.
As a result,
the allowed neutrino masses are different from those predicted in the previous
studies~\cite{Geng:2014yoa, Geng:2015vsa}.
We conclude  that the dynamical background evolution is a non-negligible effect in the viable $f(R)$ gravity theories, so that one has
 to treat the  background carefully when  $f(R)$ gravity  is investigated.
Additionally,  our conclusion  may occur in other kind of modified gravity theories, such as the Horndeski models,
which should be studied.

Finally, we have two remarks.
First, the model parameters in the Starobinsky model as well as the other power law type models have been strongly constrained by the observations
at the scale deep inside the horizon, i.e., the nonlinear cosmological structure region~\cite{Lombriser:2014dua, Cataneo:2014kaa, Liu:2016xes},
but fortunately the exponential gravity approaches the $\Lambda$CDM model rapidly when $R \gg R_{ch}$, 
with which this type of the models escapes from the fifth force problem, even if $\lambda_e \sim \mathcal{O}(1)$.
 Second, it should be emphasized  that our results of the viable $f(R)$ models based on the linear perturbation work well only 
 for $k \lesssim 0.1 h Mpc^{-1}$, whereas those for a large value of $k$ of the non-linear regime may 
 not be trustable~\cite{Oyaizu:2008tb,Hojjati:2011ix,Baldi:2013iza}.
 Clearly, the dynamical background effect with the non-linear perturbation theory should be throughly investigated in the future.

\begin{acknowledgments}
The work was supported in part by National Center for Theoretical Sciences, MoST (MoST-104-2112-M-007-003-MY3
and MoST-107-2119-M-007-013-MY3), NSFC (11547008)  
 and the Newton International Fellowship (NF160058) from the Royal Society (UK).
\end{acknowledgments}

\end{document}